\documentclass[a4paper]{article}
\pdfoutput=1

\usepackage{INTERSPEECH2022}
\usepackage{caption}
\usepackage{subcaption}
\usepackage{pifont}
\usepackage{makecell}
\usepackage{footnote}
\usepackage{threeparttable}

\usepackage[natbib, bibencoding=utf8, citestyle=numeric, bibstyle=ieee, maxbibnames=999, maxcitenames=2, mincitenames=1, sortcites]{biblatex}
\bibliography{mybib}

\usepackage[acronym, shortcuts, nohypertypes={acronym}]{glossaries}
\newacronym{DL}{DL}{deep learning}

\usepackage[activate]{microtype}
\newcommand{\ie}{i.\,e.\,}
\newcommand{\eg}{e.\,g.\,}
\newcommand{\cf}{{cf.\ }}

\usepackage[capitalise, nameinlink, noabbrev]{cleveref}

\title{An Overview \& Analysis of Sequence-to-Sequence Emotional Voice Conversion}
\name{Zijiang Yang$^1$, Xin Jing$^1$, Andreas Triantafyllopoulos$^1$, \\ Meishu Song$^1$, Ilhan Aslan$^2$, Bj{\"o}rn W. Schuller$^{1,3}$}
\address{
 $^1$Chair of Embedded Intelligence for Health Care and Wellbeing, University of Augsburg, Germany\\
 $^2$Device Software Lab, Munich Research Center, Huawei Technologies, Germany\\
 $^3$GLAM -- the Group on Language, Audio, {\&} Music, Imperial College London, UK}
\email{zijiang.yang@ieee.org}

\begin{document}

\maketitle
\begin{abstract}
Emotional voice conversion (EVC) focuses on converting a speech utterance from a source to a target emotion; it can thus be a key enabling technology for human-computer interaction applications and beyond. 
However, EVC remains an unsolved research problem with several challenges.
In particular, as speech rate and rhythm are two key factors of emotional conversion, models have to generate output sequences of differing length.
Sequence-to-sequence modelling is recently emerging as a competitive paradigm for models that can overcome those challenges.
In an attempt to stimulate further research in this promising new direction, recent sequence-to-sequence EVC papers were systematically investigated and reviewed from six perspectives: their motivation, training strategies, model architectures, datasets, model inputs, and evaluation methods.
This information is organised to provide the research community with an easily digestible overview of the current state-of-the-art.
Finally, we discuss existing challenges of sequence-to-sequence EVC.
\end{abstract}

\section{Introduction}
Speech technology has come a long way in the quest to enable human-like interactions with machines, with research increasingly addressing challenges, originally introduced in the field of affective computing~\cite{bSchuller19-ABC}. However, while humans easily convey and react to emotions in interpersonal conversations, today's machines still struggle to synthesise basic emotional speech. 

An emotional text-to-speech (ETTS), emotional voice synthesis (EVS), and emotional voice conversion (EVC) would provide user experience designers a powerful new tool to manage and navigate the challenging emotional context in conversational speech interfaces with humans~\cite{dDeibel21-CTU}. Ultimately, emotional speech generation and conversion is able to drastically change the value of many products with speech interfaces.
ETTS or EVS~\cite{oKwon19-AES, hChoi19-MEA, oKwon19-ESS} aim to \emph{directly synthesise} speech with emotional expressivity \emph{from text} -- which is a key aspect of text-to-speech (TTS) naturalness that is currently missing. 
Meanwhile, EVC~\cite{kZhou22-EVC} aims to \emph{convert} the state of a spoken utterance from one emotion to another, while preserving the linguistic information and speaker identity.
One could argue that EVC takes a `shortcut' compared to ETTS, since it endows an existing utterance with emotional intonation, rather than synthesising one from the ground up.
This allows EVC to be added as an extra step in a traditional TTS pipeline -- first synthesise, and then convert to the target emotion.
This decomposition into two constituents can reduce computational complexity and dependence on data.

Previous surveys have largely concentrated on ETTS~\cite{hGunes12-ERA, mSchroder01-ESS}; however, these are mostly outdated as they refer to literature prior to the significant milestones achieved by neural speech synthesis.
Moreover, recent surveys that incorporate the \ac{DL} paradigm have been targeted to `standard' TTS (\ie without emotional information)~\cite{xTan21-ASN}.
The most relevant survey is that of \cite{kZhou22-EVC}, which provides a comprehensive overview of EVC.
However, it devotes little attention to sequence-to-sequence (seq2seq) models, and largely concentrates on generative adversarial networks (GANs) or spectrum and prosody mapping techniques.
Thus, it leaves a small --but noteworthy-- gap of \ac{DL}-based approaches for EVC that this contribution attempts to fill.

Frame-to-frame spectral mapping is the mainstream in previous studies~\cite{zLuo16-EVC, gRizos20-SES, kZhou20-CAE}, however, emotion is inherently supra-segmental and complex with multiple signal attributes concerning both the spectrum and prosody. 
Thus, frame-based mapping of spectral features of the source and target is insufficient to convert the emotion. 
Recently, the seq2seq speech synthesis framework raises much interests in EVC. 

This research field has recently benefited from the advent of new machine learning techniques such as deep neural networks. 
Therefore, this paper aims to give an overview of recent developments, pointing out the inherent properties of the various synthesis techniques used, summarising the prosody rules employed, and analysing the evaluation paradigms. 
Finally, an attempt is made to discuss the existing challenges in EVC.

\section{Overview}
Seq2seq learning was initially proposed for machine translation by \citet{iSutskever14-SSL} and has since proved its competitiveness in several natural language processing tasks~\cite{rLiu20-VCT, cChiu18-SSR, yWang17-TTE}.
Seq2seq models consist of two main modules: the encoder and the decoder. 
Unlike the decoder taking the output of the encoder like in a standard autoencoder model~\cite{mElgaar20-MME}, seq2seq models generate the prediction of the timestep $t$ by using the prediction of the timestep $t-1$ as the input of the decoder~\cite{iSutskever14-SSL}. 
Therefore, seq2seq models are able to generate outputs with different, variable length. 
Conventional EVC, from the basic neural network architecture~\cite{zLuo16-EVC} to the recent research with GANs~\cite{gRizos20-SES, kZhou20-TSP, kZhou20-CAE}, applied frame-to-frame conversion on the spectrum and prosody, which indicates that the converted speech has the same length with the source speech. 
However, one of the vital characteristic of emotion expression is speech rate~\cite{fEyben16-GMA}, which cannot be expressed within the fixed length obtained by using the frame-to-frame EVC~\cite{hChoi21-SEV}. 
Meanwhile, the dependency between the spectrum and prosody leads to respective conversion mistakes, such as the mismatch when doing a separate study on them~\cite{kZhou22-EII}. 
Furthermore, the emotional expression in an utterance often shows only on part of the utterance (\eg in only some, but not all, words).
Seq2seq models can naturally handle this requirement by adding attention~\cite{dBahdanau15-NMT}, which makes it possible to focus on only those relevant parts~\cite{kZhou21-LDE}. 
For these reasons, seq2seq models show promising performance compared to conventional methods.

To summarise, there are three main advantages of seq2seq EVC models:

\begin{enumerate}
 \item Seq2seq models are able to learn feature mapping, alignment, and duration prediction simultaneously;
 \item Seq2seq models avoid mistakes caused by the respective mapping of spectrum and prosody;
 \item The attention mechanism helps seq2seq models focus on the emotionally emphasised parts in the utterance.
\end{enumerate}

On the other hand, seq2seq training always requires a large-sized dataset~\cite{wHuang20-VTN}. 
Moreover, the training data should be parallel, which means the same content should be expressed in different emotion categories by the same speaker -- a requirement not necessary for other models, like GANs~\citep{kZhou20-CAE, kZhou20-TSP}. 
This represents the major challenge for such models, which has so far prevented the seq2seq paradigm from becoming the dominant one in the field of EVC.
As a result, at the time of writing, only 6 papers exist which use seq2seq EVC~\cite{tKim20-EVC, cRobinson19-SMF, kZhou22-EII, zZhao21-IMS, hChoi21-SEV, kZhou21-LDE}.
These works constitute the background material for this review, and will be comprehensively analysed in the following sections.
Table~\ref{tab_evc} contains the important information in a condensed form.

\subsection{Motivation}
\citet{cRobinson19-SMF} were the first to introduce a seq2seq model to the EVC task.
They presented a model based on converting F0 of the speech. 
A three-step procedure, including F0 extraction, transformation, and subsequent application of the resulting contour on the signal, is able to convert any neutral speech to the speech with one specific emotional category. 
In other words, this is a one-to-one neutral-to-emotional EVC model.

\citet{tKim20-EVC} addressed a major problem plaguing both VC and EVC tasks -- mispronunciation.
Instead of text-supervision~\cite{jZhang19-ISV}, TTS was introduced to seq2seq EVC for guiding the linguistic information~\cite{tKim20-EVC}. 
Furthermore, this was the first many-to-many EVC system based on a seq2seq mechanism, which was facilitated by feeding a reference speech with the target emotion.

A one-to-many seq2seq VC model was presented by \citet{zZhao21-IMS}.
The authors focused on training efficiency and stability by manually balancing the word distribution and increasing the proportion of uncommon words in the dataset. 
In this way, the size of the training dataset could be also decreased to achieve a similar or even better performance. 
With the implementation of an emotion encoder, the model is able to convert high-quality emotional speech.

Considering the fact that it is impractical to find a large-sized parallel emotional dataset suitable for seq2seq training, \citet{kZhou21-LDE} presented in their recent paper a training strategy called `two-stage training', including style initialisation with a TTS dataset and emotion training. This is able to help the many-to-many EVC model improve its performance by using only a small-sized parallel emotional dataset.

Finally, the last two papers focus on emotional intensity control. 
A key difference is that \citet{hChoi21-SEV} required a parallel multi-speaker emotional dataset with the help of a speaker encoder, while the most recent solution in  \citet{kZhou22-EII} only requires a small-sized parallel single-speaker emotional dataset. 
However, \citet{hChoi21-SEV} used a weight to control the emotional intensity by multiplying it with the emotion embedding, while \citet{kZhou22-EII} trained the model with variations of intensity without any annotation on it (\cf \cref{ssec:architectures}).

\subsection{Training Strategies}
\label{sec:strategies}
One of the most common training strategies utilised for seq2seq models is called \emph{teacher forcing}.
In seq2seq training, the \textit{generated} frame of the previous timestep will be fed into the decoder to generate the frame of the current timestep (using the pre-defined start of the sentence token to generate the frame of the first timestep)~\cite{iSutskever14-SSL}. However, this causes a problem when training, because the error will be accumulated during generating. 
Moreover, the generating takes a long time, since the generating is frame-by-frame. 
Teacher forcing feeds the \textit{ground truth} frame instead of the generated frame in the training phase. 
Therefore, it has the ability to help the model learn faster and more accurately, especially at the beginning of the training~\cite{cChiu18-SSR}.

However, specific challenges arising in EVC require specialised training regiments.
In order to solve mispronunciation and training instability without explicit alignment mechanisms, \citet{tKim20-EVC} applied multi-task learning by introducing TTS to the EVC task. Besides the content encoder which generates linguistic embedding by using the source speech, a text encoder was implemented to encode the input text to a linguistic embedding.
Then, during training, the model was randomly tasked to perform either EVC or TTS -- a form of alternating multi-task learning which helped it avoid mispronunciation errors.

Since it is impractical to use a large-sized parallel dataset to fulfil the requirements of seq2seq EVC training, \citet{kZhou21-LDE} proposed a two-stage training strategy, which begins with a style initialisation phase with a large-sized TTS corpus before doing emotional fine-tuning on a small-sized emotional dataset. 
Moreover, an emotion classifier was utilised in adversarial fashion to eliminate the emotional information in the linguistic embedding~\cite{kZhou21-LDE}. 
This adversarial training strategy was aimed to optimise the performance on disentangling the style/emotional information and the linguistic information. 
\citet{zZhao21-IMS} presented a similar work by utilising an emotional embedding to the pre-trained VC model.

\citet{kZhou22-EII} improved their model by adding emotion supervision training with a pre-trained speech emotion recognition (SER) module.
Accounting for the fact that the reconstruction loss between the target speech and the converted speech does not incorporate human emotional perception, a SER module was introduced to compute two perceptual losses: emotion classification loss and emotion embedding similarity loss, thus optimising the emotional perception of the converted speech.

\begin{table*}[th]
 \caption{Information of all sequence-to-sequence EVC papers. \textbf{Abbreviations}: \textbf{ABX}: ABX test, \textbf{BWS}: Best-Worst Scaling, \textbf{CER}: Character Error Rate, \textbf{CS}: Cosine Similarity, \textbf{CTC}: Connectionist Temporal Classification, \textbf{DDUR}: Differences of Duration, \textbf{FFE}: F0 Frame Error, \textbf{GPE}: Gross Pitch Error, \textbf{MCD}: Mel-cepstral Distortion, \textbf{MOS}: Mean Opinion Score, \textbf{SSER}: Subjective Speech Emotion Recognition, \textbf{VDE}: Voicing Decision Error, \textbf{WER}: Word Error Rate.} 
 
 \label{tab_evc}
 \centering
 \resizebox{\textwidth}{!}{
 
\begin{threeparttable}
 \begin{tabular}{c c c c c c c c}
 \toprule
 
  \textbf{Paper} &
  \textbf{Highlights} &
  \makecell{\textbf{Feature Set} \\ \textbf{\& Vocoder}} & 
  \makecell{\textbf{Emotional} \\ \textbf{Dataset}} &
  \textbf{Language} & 
  \makecell{\textbf{Emotional} \\ \textbf{Model}} & 
  \makecell{\textbf{Evaluation} \\ \textbf{Methods}} &
  \makecell{\textbf{Public} \\ \textbf{Code}} \\
  
 \midrule
  \cite{cRobinson19-SMF} &
  \makecell{First work \\ Syllable-level conversion} & 
  \makecell{F0 \\ SuperVP} & 
  \makecell{$\mathtt{\sim}$1\,100 \\ syllables} & 
  \makecell{ \\ French \\ \,} & 
  One-to-one & 
  SSER & 
  \ding{51}$^1$ \\
 \midrule
 
  \cite{tKim20-EVC} &
  Multi-task learning &
  \makecell{Log Mel-spectrogram \\ Griffin-Lim Algorithm} &
  \makecell{21\,000 \\ utterances} &
  \makecell{ \\ Korean \\ \,} & 
  Many-to-many & 
  \makecell{WER CS \\ MOS ABX} & 
  \ding{51}$^2$ \\
  
 \midrule
 
  \cite{zZhao21-IMS} &
  \makecell{Data redundancy reduction \\ CTC leverage \\ EVC fine-tuning} & 
  \makecell{Log Mel-spectrogram \\ HiFi-GAN} & 
  \makecell{6\,000 \\ utterances} & 
  Chinese & 
  One-to-many & 
  \makecell{WER CER \\ MOS} & 
  \ding{55} \\
  
 \midrule
  
  \cite{hChoi21-SEV} &
  \makecell{Multi-speaker emo. dataset \\ Context preservation \\ Emotional intensity} & 
  \makecell{Log Mel-spectrogram \\ Parallel WaveGAN} & 
  \makecell{4\,000 \\ utterances} & 
  Korean & 
  One-to-many &
  \makecell{MCD VDE \\ GPE FFE \\ MOS ABX SSER}& 
  \ding{55} \\
  
 \midrule
 
  \cite{kZhou21-LDE} &
  \makecell{Two-stage training \\ Small emo. dataset} & 
  \makecell{Log Mel-spectrogram \\ WaveRNN} & 
  \makecell{350 \\ utterances} & 
  \makecell{ \\ English \\ \,} & 
  Many-to-many &
  \makecell{MCD DDUR \\ MOS BWS} &
  \ding{51}$^3$ \\
  
 \midrule
 
  \cite{kZhou22-EII} &
  \makecell{Style-pretraining \\ Emotion supervision training \\ Small emo. dataset \\ Emotional intensity} & 
  \makecell{Log Mel-spectrogram \\ Parallel WaveGAN} & 
  \makecell{350 \\ utterances} & 
  English & 
  Many-to-Many &
  \makecell{MCD DDUR \\ MOS BWS} &
  \ding{51}$^4$ \\
  
 \bottomrule
 \end{tabular}
 
\begin{tablenotes}
    \item $^1$ \url{https://github.com/carl-robinson/voice-emotion-seq2seq} \\ $^2$ \url{https://github.com/ktho22/vctts} \\ 
    $^3$ \url{https://github.com/KunZhou9646/seq2seq-EVC} \\ 
    $^4$ \url{https://github.com/KunZhou9646/Emovox}
\end{tablenotes}
\end{threeparttable}
}
\end{table*}

\subsection{Model Architectures}
\label{ssec:architectures}
As the first to explore seq2seq EVC, \citet{cRobinson19-SMF} used the simplest model architecture, comprising one encoder, one decoder, and one attention module. 
The encoder accepts the extracted features as the input and generates the context vector, while the decoder uses this vector and the previous frames to generate the converted features frame by frame. 
The attention mechanism is used to provide an explicit alignment between the input (source) and the output (converted).

Compared to the basic architecture above, \citet{tKim20-EVC} modified it on the encoder by using three encoders instead of one: a style, a content, and a text one. 
In the training phase, reference speech was sent to the style encoder for the style embedding, while the contents encoder and the text encoder generated the linguistic embedding from the source speech and the input text, respectively. 
Then, the style embedding along with the linguistic embedding from either the speech or the text were fed into the decoder to construct the converted emotional speech, in a multi-task way (\cf \cref{sec:strategies}). 
In the end, this model has the ability to perform both an EVC task (without the text encoder) and an emotional TTS task (without the content encoder).

\citet{kZhou21-LDE} also used three encoders: a style/emotion encoder for style/emotion information, seq2seq automatic speech recognition (ASR) utilised for linguistic information from speech features, and a text encoder which is also used to capture for linguistic information but from the input text instead. 
Furthermore, an emotion classifier was applied to optimise the linguistic embedding obtained from the source speech.

Following up on this, \citet{kZhou22-EII} added two extra modules to control the emotional intensity and improve the emotional expressivity of the output speech. 
Based on the assumption that the emotional intensity can be regarded as the relative difference from the neutral speech (zero intensity) to the emotional speech, relative attributes were applied to train the emotional intensity modelling without any explicit labels.
Subsequently, the intensity embedding, which can be derived from the reference speech or given manually, was concatenated with the emotion embedding and the resulting embedding was fed into the decoder to reconstruct the emotional speech with the required intensity. 
Furthermore, they added a pre-trained SER model and used it to generate two perceptual losses to improve performance. 
An emotion classification loss was computed by the converted emotional speech being classified by the SER model and compared with the ground truth emotional category, whereas an emotion embedding similarity loss was computed by using the emotion embedding obtained from the emotion encoder and the SER embedding obtained by sending the converted speech to the SER model. 
Visualised results showed the perceptual losses helped the emotion encoder discriminate the different emotion categories.

Finally, \citet{hChoi21-SEV} and \citet{zZhao21-IMS} both applied a speaker encoder, which is used for disentangling the speaker information and makes the use of a multi-speaker emotional dataset possible.
Moreover, \citet{hChoi21-SEV} implemented one source decoder and one target decoder in the training phase, to make sure that the content embedding of the source and the target speech preserve the contextual information by comparing the output of these decoders with the source and the target speech, respectively. 
On the other hand, \citet{zZhao21-IMS} applied a length regulator module for the length alignment between the encoders and the decoder, and used a connectionist temporal classification (CTC) recogniser~\cite{sKim17-JCB} after the decoder to guide the alignment between the text and the speech to improve the performance of the EVC model.

\subsection{Datasets}
To achieve a decent performance by using seq2seq training, a large-sized dataset is very essential~\cite{wHuang20-VTN}. Specifically, seq2seq EVC training requires a large-sized, parallel, one-speaker, emotional dataset. For instance, \citet{tKim20-EVC} utilised a Korean emotional dataset (mKETTS) including 3\,000 utterances per emotional category pronounced by one male speaker, and there are 7 different emotions in total (\textit{neutral}, \textit{anger}, \textit{disgust}, \textit{fear}, \textit{happiness}, \textit{sadness}, and \textit{surprise}). In \citet{cRobinson19-SMF}, the dataset used contains only 200 emotional utterances (10 sentences~$\times$~4 emotional category~$\times$~5 levels of intensity, \textit{anger}, \textit{joy}, \textit{fear}, and \textit{sadness}) recorded by one French actress. However, the researched conversion in this work was on the syllable-level. Thus, the model was trained by using around 1\,100 syllable pairs after forced alignment~\cite{cVeaux11-ICN}.

A speaker encoder allows the use of a multi-speaker dataset -- a method utilised by both \citet{zZhao21-IMS} and \citet{hChoi21-SEV}.
\citet{zZhao21-IMS} used a Chinese emotional dataset including three speakers, three emotional categories (\textit{anger}, \textit{happiness}, and \textit{sadness}) and 6 hours of recording to fine-tune their pre-trained VC model. Similarly, the dataset in \citet{hChoi21-SEV} includes 100 sentences in 4 different emotional categories (\textit{neutral}, \textit{anger}, \textit{happiness}, and \textit{sadness}) pronounced by 5 Korean actors and 5 Korean actresses, for a total of 4\,000 utterances. 

Instead of using a speaker encoder to expand the range of usable emotional datasets, \citet{kZhou21-LDE, kZhou22-EII} utilised two-stage training on the model to reduce its dependency on the size of the emotional dataset. At the first stage, about 30 hours of recordings recorded by 99 speakers from a multi-speaker corpus called VCTK~\cite{jYamagishi19-VCT} were applied to pre-train the model. Then, only 350 pairs of emotional speeches from ESD~\cite{kZhou22-EVC} were used to fine-tune, enabling them to improve their performance.

\subsection{Model Inputs}
Most seq2seq EVC research uses log Mel-spectrograms~\cite{zZhao21-IMS, tKim20-EVC, hChoi21-SEV, kZhou21-LDE, kZhou22-EII}, and the applied vocoders include Parallel WaveGAN~\cite{rYamamoto20-PWG}, WaveRNN~\cite{nKalchbrenner18-ENA}, and HiFi-GAN~\cite{jKong20-HFG}. Since \citet{cRobinson19-SMF} focused on F0 conversion on the syllable-level, they estimated the source F0 contour and used the converted F0 contour to synthesise the speech with the SuperVP vocoder.

An important consideration concerns the control of the target emotion.
Besides the simple one-to-one EVC~\cite{cRobinson19-SMF}, there are three different methods to bring the information of the target emotion to the model. 
The most direct method is feeding an emotion ID to the emotion encoder~\cite{zZhao21-IMS}.
In \citet{tKim20-EVC}, an emotion-reference speech was utilised in the inference phase to guide the model. 
Emotional embeddings were also used~\cite{hChoi21-SEV, kZhou21-LDE, kZhou22-EII}, where they were calculated by the average of a set of emotional speech embeddings from the same emotional category.

\subsection{Evaluation Methods}
There are two types of evaluation methods applied in this field: objective and subjective ones. In general, objective evaluation entails the calculation of some measure of difference or correlation between the output and the target. 
For example, \citet{tKim20-EVC} used word error rate (WER) for the linguistic consistency and cosine similarity for the performance on emotion conversion. 
Similarly, both WER and character error rate (CER) were applied in \cite{zZhao21-IMS}. 
\citet{kZhou21-LDE, kZhou22-EII} preferred using Mel-cepstral distortion (MCD) to measure the spectral changes during the conversion, and the differences of duration (DDUR) for the performance on the duration of the converted speech. 
Besides MCD, \citet{hChoi21-SEV} applied three other objective evaluation methods: voicing decision error (VDE), gross pitch error (GPE), and F0 frame error (FFE), for a more comprehensive analysis of their results.

Moreover, several different subjective evaluation methods were applied: \citet{cRobinson19-SMF} did a subjective emotion recognition survey with 87 participants to measure the performance of the EVC model from a human perspective. Additionally, mean opinion score (MOS)~\cite{zZhao21-IMS, hChoi21-SEV, kZhou21-LDE, kZhou22-EII}, ABX test (identifying whether sample X is from class A or B)~\cite{hChoi21-SEV}, and best worst scaling (BWS)~\cite{kZhou21-LDE, kZhou22-EII} on the naturalness, clarity, and similarity were also applied in prior works. 

\section{Challenges and Conclusion}
Although an alternative solution was proposed by \citet{kZhou21-LDE, kZhou22-EII} to alleviate the requirement on a large-sized parallel emotional dataset, the main challenge of seq2seq EVC task remains the availability of appropriate public datasets. Existing datasets are small or lacking in quality.
For example, the newly published dataset ESD~\cite{kZhou22-EVC} is a clean, parallel dataset but includes only 350 utterances per speaker. EmoV-DB~\cite{aAdigwe18-EVD} has more pairs of samples; however, there are non-speech utterances included, such as laughter and yawn. 
The collection and release of suitable datasets to the public would foster further research in this promising field and help improve the performance.

Another remaining challenge is the difficulty in comparing different EVC models.
Objective evaluation methods are not intuitive enough because they only indicate how `close' or `similar' the converted speech and the target speech are with respect to some metric -- however, this is not a guarantee that human perception will also consider them as close.
Using evaluation methods based on combinations of pre-trained SER and ASR
models, or trying to predict subjective evaluation scores --as is the target of the INTERSPEECH 2022 VoiceMOS challenge\footnote{https://voicemos-challenge-2022.github.io/}-- are both promising methods of mitigating this challenge.
However, until those methods mature enough to enable their usage for practical applications, subjective annotations will remain the gold-standard for EVC evaluations.

Nevertheless, those come with their own drawbacks, for example, when participants between studies are biased (\eg due to different cultural backgrounds). 
Using different validation sentences from different datasets is another problem, since context strongly affects the emotion that humans experience~\cite{aMetallinou12-CLE}. 
Finally, different evaluation methods on the same criteria can make it more challenging to compare. For example, \citet{hChoi21-SEV} applied MOS while \citet{kZhou22-EII} utilised BWS on the speech similarity assessment.
With improvements in EVC quality, there will be a need to move away from simple MOS to evaluate utterances which are detached from any context or interaction scenarios. New qualitative evaluation methods will be needed to identify fine grained differences between different EVC solutions, considering conversation contexts and speaker intents in more detail. To this end, inspiration can also be taken from related fields, especially the field of Human-Robot Interaction, where researchers usually evaluate the affect of robots in interactive settings, analysing both users' immediate reactions and their preferences. For example, \citet{hRitschel19-IMA} have evaluated the effect of a robot converting non-ironic utterances into ironic utterances in small talks with users in order to be more likeable.

In conclusion, seq2seq EVC is a promising, rapidly maturing research field.
While there still remain several challenges (which are not unique to this paradigm), these models have the potential to improve the performance of EVC applications, thus leading to more intelligent human-computer interactions.
We presented a short, concise review of recent approaches, which we hope to also fuel novel approaches.


\section{Acknowledgements}

This research was partially supported by the Affective Computing \& HCI Innovation Research Lab between Huawei Technologies and University of Augsburg, and the China Scholarship Council~(CSC)~, Grant \#\,202006290013.

\newpage

\section{\refname}
 \printbibliography[heading=none]



\end{document}